\documentclass[showpacs,preprintnumbers,amsmath,amssymb,aps]{revtex4}

\def\be{\begin{equation}}
\def\ee{\end{equation}}
\def\bea{\begin{eqnarray}}
\def\eea{\end{eqnarray}}
\def\A{N}
\def\H{{u}}
\def\R{{\cal{R}}}

\def\sigy{\sigma_y}
\def\sigt{\sigma_t}
\def\Att{A}
\def\Aty{A_{y}}
\def\Ayy{A_{yy}}
\def\lAtt{\tilde{A}}
\def\lAty{\tilde{A}_{y}}
\def\lAyy{\tilde{A}_{yy}}
\def\lR{\tilde{\cal{R}}}

\begin{document}

\title{Scalar perturbations from brane-world inflation}

\author{Kazuya Koyama$^{1,2}$, David Langlois$^{3,4}$, Roy Maartens$^1$
and David Wands$^1$\\~}

\affiliation{$^1$Institute of Cosmology \& Gravitation,
University of Portsmouth, Portsmouth~PO1~2EG, UK\\
$^2$Department of Physics, University of Tokyo, Tokyo~113-0033, Japan \\
$^3$GRECO (FRE~2435-CNRS), Institut d'Astrophysique de Paris,
75014~Paris, France\\
$^4$F\'ed\'eration de recherche APC, Universit\'e Paris VII,
75251~Paris, France}

\date{\today}

\begin{abstract}

We investigate the scalar metric perturbations about a de Sitter
brane universe in a 5-dimensional anti de Sitter bulk.  We compare
the master-variable formalism, describing metric perturbations in
a 5-dimensional longitudinal gauge, with results in a Gaussian
normal gauge. For a vacuum brane (with constant brane tension)
there is a continuum of normalizable Kaluza-Klein modes, with $m>
{3\over2}H$, which remain in the vacuum state. A light radion
mode, with $m=\sqrt{2}H$, satisfies the boundary conditions for
two branes but is not normalizable in the single-brane case. When
matter is introduced (as a test field) on the brane, this mode,
together with the zero-mode and an infinite ladder of discrete
tachyonic modes, become normalizable. However, the boundary
condition requires the self-consistent 4-dimensional evolution of
scalar field perturbations on the brane and the dangerous growing
modes are not excited. These normalizable discrete modes introduce
corrections at first-order to the scalar field perturbations
computed  in a slow-roll expansion.  On super-Hubble scales, the
correction is smaller than slow-roll corrections to the de Sitter
background. However on small scales the corrections can become
significant.

\end{abstract}


\maketitle

\section{Introduction}

If gravity propagates in extra spatial dimensions, while Standard
Model fields are confined to the 3 observed spatial dimensions,
then the observable universe could be described by a 3-brane in a
$4+d$-dimensional bulk spacetime
(see~\cite{Reviews1,Reviews2,Reviews3} for reviews on the
subject). At low energies, the extra-dimensional effects should be
small in order to recover the successes of 4-dimensional general
relativity, but at high energies these effects could be dominant.
If the early universe included a period of inflation, then the
extra-dimensional gravitational effects could introduce
significant changes to the dynamics and generation of primordial
perturbations at high energy. Any imprints left on the
perturbation spectra will be constrained by increasingly high
precision observations of the cosmic microwave background,
providing in principle constraints on extra-dimensional theories.

The study of general cosmological perturbations in brane-worlds is
complicated because the motion of the brane in the
higher-dimensional bulk makes it impossible, in general, to
separate the evolution of different Kaluza-Klein modes. It is only
possible to fix the brane location and obtain a separable wave
equation for perturbations in the special case of a de Sitter (or
Minkowski or anti-de Sitter) brane in 5-dimensional anti-de Sitter
(AdS) spacetime~\cite{Kaloper,GS}. This is useful as it provides a
zeroth-order approximation for slow-roll inflation on the brane.
In this case the behaviour of tensor~\cite{LMW} or
vector~\cite{BMWvector} metric perturbations has been described,
while neglecting matter sources on the brane. In particular one
can estimate high-energy corrections to the spectrum of
gravitational waves produced from vacuum fluctuations in the bulk
spacetime during inflation~\cite{LMW,Rubakov,fk,Tanaka}.

The amplitude of scalar perturbations on the brane due to inflaton
field fluctuations has also been estimated in the extreme
slow-roll limit where the coupling of field fluctuations to bulk
metric perturbations is neglected~\cite{MWBH99}. Energy
conservation on the brane is sufficient to ensure that there
exists a scalar curvature perturbation for matter that is
conserved for adiabatic density perturbations in a large scale
limit~\cite{WMLL,LMSW}. In this limit, this approach by-passes the
need to study bulk scalar metric perturbations coupled to matter
on the brane. But we do need to understand the bulk metric
perturbations in order to go beyond the zeroth-order slow-roll
approximation (for partial attempts see
Refs.~\cite{Calcagni,Liddle,Seery}) and to distinguish
5-dimensional effects from a modified 4-dimensional
theory~\cite{LiddleTaylor}. (Note that the bulk metric
perturbations are also needed in order to compute the Sachs-Wolfe
effect, since the large-scale curvature perturbation does not
determine the brane metric perturbations~\cite{LMSW,kcmb}.)

In this paper we investigate bulk scalar metric perturbations
about a de Sitter brane. We first consider scalar metric
perturbations in the absence of matter perturbations. In this case
there are no normalizable light modes (with effective
4-dimensional mass $m^2< {9\over4}H^2$) for a single
brane~\cite{fk}, but in the presence of a second brane there is a
normalizable ``radion'' mode~\cite{GenSasaki}. We discuss how this
discrete mode appears either as a displacement (``bending") of the
brane or as a bulk metric perturbation in different gauges. The
radion is massless for two Minkowski branes~\cite{GarrigaTanaka}
but appears as an ``instability'' for two de Sitter
branes~\cite{GenSasaki,ChackoFox,BDLradion,ckp}, although the
effect of the radion ``instability'' on the brane becomes small on
large scales~\cite{GenSasaki2}.

We then go on to consider the bulk metric perturbations excited by
scalar field fluctuations on a single de Sitter brane as a first
step towards calculating the effect of first-order slow-roll
corrections on the exact de Sitter solutions. We show that scalar
field perturbations can excite an infinite ladder of apparently
tachyonic, normalizable bulk modes. However the boundary condition
requires the self-consistent 4-dimensional evolution of scalar
field perturbations on the brane and the dangerous growing modes,
that one might expect to find for tachyonic modes, are not
allowed. We comment on the possible effect of metric backreaction
upon the scalar fluctuations during inflation.

We present our conclusions in Section~\ref{conc}.

\section{Randall-Sundrum Cosmology}

The Randall-Sundrum (RS) model~\cite{RS99} provides the basis for
a simple realization of the brane-world idea in
cosmology~\cite{BDL,SMS,BDEL}. The background bulk is 5D AdS
spacetime with a negative cosmological constant $\Lambda_5$ and
the brane has Friedman-Robertson-Walker geometry. For a general
brane and bulk geometry, the 5D field equations are
\begin{equation}
 \label{5DEinstein}
{}^{(5)\!}G_{AB} + \Lambda_5\, {}^{(5)\!}g_{AB} = 0 \,.
\end{equation}
One can define an energy scale $\mu$ corresponding to the
curvature scale of the bulk, via $\Lambda_5=-6\mu^2$.

The induced metric on the brane is
\begin{equation}
g_{AB} = {}^{(5)\!}g_{AB} - n_{A}n_{B}\,,
\end{equation}
where $n^A$ is the unit vector normal to the brane. The 4D matter
fields determine the brane trajectory in the bulk via the junction
conditions, by producing the jump in the extrinsic curvature at
the brane.  Without loss of generality, the surface
energy-momentum on the brane can be split into two parts,
$T_{\mu\nu}-\lambda g_{\mu\nu}$, where $T_{\mu\nu}$ is the matter
energy-momentum tensor and $\lambda$ is a constant brane tension.
The junction condition with $Z_2$-symmetry is then~\cite{BDL,SMS}
\begin{equation}
\label{Israel} K^{\mu~+}_\nu - K^{\mu~-}_\nu = 2K^{\mu~+}_\nu =
-\kappa_5^{2} \left[ T^\mu_\nu-{1\over 3} \delta^\mu_\nu \left( T
- \lambda \right) \right],
\end{equation}
where the extrinsic curvature of the brane is $K_{\mu\nu}=
g_\mu^{~C} g_\nu^{~D} \left[{}^{(5)\!}\nabla_C n_D\right]$ and
$\kappa_5^{2}$ is the 5-dimensional coupling of matter to gravity.
The effective Einstein equations for the induced metric on the
brane are then~\cite{SMS}
\begin{equation}
\label{EE} G_{\mu\nu} = \kappa_4^2 T_{\mu\nu} + \kappa_5^4
\Pi_{\mu\nu} - E_{\mu\nu} \,,
\end{equation}
where the effective 4D coupling of matter to gravity on the brane
at low energies is given by $\kappa_4^2=\mu\kappa_5^2$ and we have
chosen the arbitrary constant $\lambda=6\mu/\kappa_5^2$. As well
as the high-energy corrections due to the tensor $\Pi_{\mu\nu}$
which is quadratic in the energy-momentum tensor $T_{\mu\nu}$, the
effective Einstein equations include a non-local contribution
$E_{\mu\nu}$ from the projection of the 5D Weyl tensor.

In order to study inhomogeneous bulk metric perturbations, we
choose a specific form for the unperturbed 5D spacetime that
accommodates any spatially flat FRW cosmological solution on the
brane at $y=0$,
\begin{equation}
\label{backmetric} ds^2 = - n^2(t,y) dt^2 + a^2(t,y) d\vec{x}\,^2
+ b^2(t,y) dy^2 \,.
\end{equation}
The scale factor on the brane is $a_o(t)=a(t,0)$. The junction
conditions~(\ref{Israel}) for this background metric yield
\begin{equation}
\label{junction_back}
\frac{a'^+-a'^-}{a}=-\frac{\kappa_5^{2}}{3}\left(\lambda+\rho\right),
\qquad \frac{n'^+-n'^-}{n}=-\frac{\kappa_5^{2}}{3}\left(\lambda-3P
-2\rho\right),
\end{equation}
where $\rho$ and $P$ are respectively the energy density and
pressure associated with the homogeneous brane energy-momentum
tensor $T_{\mu\nu}$.

Allowing arbitrary first-order scalar metric perturbations then
gives the metric~\cite{BMW,Deffayet}
\begin{equation}
\label{pertmetric} g_{AB}= \left[
\begin{array}{ccc}
-n^2(1+2\Att) & a^2B_{,i} & n\Aty \\
a^2B_{,j} & a^2\left\{ (1+2\R)\delta_{ij} + 2E_{,ij} \right\} &
a^2B_{y,i} \\
n\Aty & a^2B_{y,i} & b^2(1+2\Ayy)
\end{array}
\right] \,.
\end{equation}
Note that we are using the common cosmological notation of scalar
perturbations to denote scalars with respect to 3-space slices at
fixed $t$ and $y$.

The perturbed energy-momentum tensor for matter on the brane, with
background energy density $\rho$ and pressure $P$, can be written
as
\begin{equation}
\label{Tmunu} T^\mu_\nu = \left[
\begin{array}{ccc}
-(\rho+\delta\rho) & & \delta q_{,j} \\ &&\\
-a^{-2}\left\{\delta q^{,i}-(\rho+P)B^{,i}\right\} & & (P+\delta
P)\delta^i_j + \delta\pi^i_j
\end{array}
\right] \,,
\end{equation}
where $\delta\pi^i_j=\delta\pi^{,i}{}_{,j}-{1\over3}
\delta^i_j\,\delta\pi^{,k}{}_{,k}$ is the tracefree anisotropic
stress perturbation. Substituting in the junction conditions, this
requires (see e.g.~\cite{L00a,L00b}),
\begin{eqnarray}
\delta K_0^0 &=& \frac{\kappa_5^2}{6} \left( 2\delta\rho + 3\delta
P \right) \,, \label{perturbjunction}
\\
\delta K_i^0 &=& - \frac{\kappa_5^2}{2} \delta q_{,i} ~,\\
\delta K_j^i &=& - \frac{\kappa_5^2}{6} (\delta\rho
 -\vec{\nabla}^2 \delta \pi) \delta_j^i
-\frac{\kappa_5^2}{2} \delta \pi^{,i}{}_{,j} ~.
\label{perturbjuncTT}
\end{eqnarray}
The components of the perturbed extrinsic curvature in an
arbitrary gauge are given by~\cite{MRL}
\begin{eqnarray}
\delta K^{0}_0 &=& \frac{1}{b} \left[A' - \frac{n'}{n}A_{yy} +
\frac{1}{n} \dot{A}_y + \frac{b^2}{n^2} \left\{\ddot{\xi} +
\left(2 \frac{\dot{b}}{b} - \frac{\dot{n}}{n} \right) \xi \right
\} + \left\{\left( \frac{n'}{n} \right)'- \frac{n'}{n}
\frac{b'}{b}
\right\} \xi \right], \\
\delta K^i_j &=& \frac{1}{b} \left[ \R' - \frac{a'}{a} A_{yy} +
\frac{1}{n^2} \frac{\dot{a}}{a} (n A_y + b^2 \dot{\xi}) + \left\{
\left(\frac{a'}{a}\right)' - \frac{a'}{a} \frac{b'}{b}  \right \}
\xi
\right] \delta^{i}_j \nonumber\\
&& + \frac{1}{b} \left[E'-B_y - \frac{b^2}{a^2} \xi
\right]^{,i}_{~,j}~, \label{perturbKTT}
\\
\delta K^{0}_i &=& -n^{-2} \left[ \frac{1}{2} \frac{a^2}{b}
\left(B' - \dot{B}_y -\frac{n}{a^2} A_y \right) +b \left\{ -
\dot{\xi}+ \left(\frac{\dot{a}}{a} - \frac{\dot{b}}{b} \right) \xi
\right\} \right]_{,i}\, , \label{perturbK}
\end{eqnarray}
where a dot denotes a derivative with respect to time $t$ and a
prime a derivative with respect to $y$. Here we took into account
the fact that the position of the brane is generally displaced
from $y=0$ in a general gauge. The brane bending scalar
$\xi(t,\vec{x})$ describes the perturbed position of the brane.

\section{Master variable and alternative bulk  gauges}

\subsection{5D longitudinal gauge}
\label{S5Dlong}

To eliminate any gauge dependence on the choice of 3-space
coordinates we can work with the spatially gauge-invariant
combinations
\begin{equation}
 \sigt = -B + \dot{E} \, ,~~
 \sigy = -B_y + E' \, ,
\label{scalarGI}
\end{equation}
which are subject only to temporal and bulk gauge transformations.
The bulk and temporal gauges are fully determined by setting
$\sigt=\sigy=0$, which we refer to as the 5D longitudinal
gauge~\cite{cvdb,BMW} to avoid possible confusion with quantities
in the 4D longitudinal gauge on the brane.

We can define the remaining metric perturbations in the 5D
longitudinal gauge as
\bea \label{5Dlongpert} \lAtt &=& \Att- \frac{1}{n}
\left(\frac{a^2}{n^2}\sigt\right)^{\displaystyle{\cdot}}
 + \frac{n'}{n} \frac{a^2}{b^2} \sigy \, ,  \\
\lR &=& \R -\frac{\dot a}{a} \frac{a^2}{n^2} \sigt + \frac{a'}{a}
\frac{a^2}{b^2} \sigy \, ,  \\
\lAty &=& \Aty +n\left(\frac{a^2}{n^2}\sigt\right)'
+\frac{b^2}{n}\left( \frac{a^2}{b^2}
\sigy\right)^{\displaystyle{\cdot}} \, ,
 \\
\lAyy &=& \Ayy - \frac{\dot{b}}{b} \frac{a^2}{n^2} \sigt +
\frac{1}{b} \left( \frac{a^2}{b} \sigy\right)' \, .
\eea
These are equivalent to the gauge-invariant bulk perturbations
originally introduced in covariant form in~\cite{Mukoh,Kodama} and
in a coordinate-based approach in~\cite{cvdb}. The spatial trace
part of the 5D Einstein equations simplifies in the 5D
longitudinal gauge to
\begin{equation}
\lAtt + \lR + \lAyy=0 \,.
\end{equation}

Mukohyama~\cite{Mukoh} (see also~\cite{Kodama}) was the first to
show that the perturbed 5D Einstein equations, in the absence of
bulk matter perturbations, $^{(5)\!}\delta G^A_B=0$, are solved in
an AdS background if the metric perturbations are derived from a
``master variable'', $\Omega$:
\bea
 \label{lA}
 \lAtt &=& -\frac{1}{6a}
 \left\{ \frac{1}{b^2} \left[
  2\Omega'' - \left( 2\frac{b'}{b}+\frac{n'}{n}\right) \Omega'
  \right]
+ \frac{1}{n^2} \left[ \ddot\Omega - \left(
2\frac{\dot{b}}{b}+\frac{\dot n}{n} \right) \dot\Omega \right]
  - \mu^2 \Omega
 \right\}
\, , \\
\label{lAy} \lAty &=& \frac{1}{na}\left(
 \dot\Omega'-\frac{n'}{n}\dot\Omega - \frac{\dot{b}}{b}\Omega'
  \right)\,, \\
\label{lAyy}
 \lAyy &=& \frac{1}{6a} \left\{
  \frac{1}{b^2} \left[
\Omega''- \left( 2\frac{n'}{n} + \frac{b'}{b} \right) \Omega'
\right]
 + \frac{1}{n^2} \left[
2\ddot\Omega-\left( 2\frac{\dot{n}}{n} + \frac{\dot{b}}{b} \right)
  \dot\Omega\right]
 +\mu^2 \Omega \right\} \,, \\
\label{lR} \lR &=&
 \frac{1}{6a}\left\{
 \frac{1}{b^2} \left[
\Omega''+ \left(\frac{n'}{n}-\frac{b'}{b}\right) \Omega' \right]
 +\frac{1}{n^2}\left[
-\ddot\Omega + \left( \frac{\dot n}{n}-\frac{\dot{b}}{b}\right)
\dot\Omega
  \right]
 - 2\mu^2 \Omega\right\} \,.
\eea
The remaining perturbed 5D Einstein equations then yield a single
wave equation governing the evolution of the master variable
$\Omega$ in the bulk:
\begin{equation}
\label{scalarmastereom}
 - \left( {b\over na^3} \dot\Omega \right)^{\displaystyle\cdot}
 + \left( {n\over b a^3} \Omega' \right)^\prime
 + \left( \mu^2 - {k^2\over a^2} \right)
 {nb\over a^3} \Omega
 = 0
 \,,
\end{equation}
where $k$ is the comoving wave-number along the brane,
$\vec{\nabla}^2\to -k^2$. Note that this is not the standard form
for a 5-dimensional wave-equation for a canonical scalar field. It
can be re-written in a standard form by defining $\omega \equiv
a^{-3} \Omega$ but we shall work with the original variable
introduced in~\cite{Mukoh}.

The advantage of the master variable approach is that the 5D field
$\Omega$ describes all the degrees of freedom of the bulk scalar
metric perturbations. In particular the perturbed brane location
$\xi$ is directly related to the anisotropic stress by boundary
conditions at the brane [see Eqs.~(\ref{perturbjuncTT}) and
(\ref{perturbKTT})]. Hence any radion mode describing the
perturbation in the relative distance between two branes must be
encoded in the bulk metric perturbations.

\subsection{Gaussian normal gauge}

An alternative choice of gauge commonly used is a Gaussian normal
(GN) gauge, where the bulk $y$ coordinate measures the proper
distance in the bulk, including first-order metric
perturbations~\cite{BMW}. This requires the metric perturbations
$B_y$, $\Aty$ and $\Ayy$ to vanish, but leaves a residual gauge
freedom to pick the 4D gauge on any given constant-$y$
hypersurface (analogous to the residual gauge freedom on a spatial
hypersurface in the 4D synchronous gauge).

At this stage, one must make a distinction between a {\it general
Gaussian Normal gauge} defined as above and the particular {\it
brane Gaussian Normal gauge} in which the above conditions are
supplemented by the requirement that the brane lies at $y=0$.

A technique used to study the propagation of gravitational waves
in a vacuum spacetime is to work in a (general) GN gauge in which
the perturbations are transverse and tracefree in the background
spacetime. The transverse and tracefree condition in a Gaussian
normal gauge actually over constrains the problem except for the
special case of a maximally symmetric 4D (anti-)de Sitter
brane~\cite{BMW}.

When the 5D perturbations of the metric $h_{AB}\equiv \delta
{}^{(5)}g_{AB}$ are transverse, $^{(5)\!}\nabla^C\, h_{AC}=0$, and
tracefree, $^{(5)\!}g^{AB}\,h_{AB}=0$,  the perturbed Einstein
equations can be written as a wave equation,
\begin{equation}
^{(5)\!}\Box\, h_{AB}
 = 2 \, ^{(5)\!}R_{CADB} \,h^{CD} \,,
\end{equation}
where $^{(5)\!}\Box=\,^{(5)\!}\nabla_C\,^{(5)\!}\nabla^C$. The
background Riemann tensor in AdS$_5$ is given by
\begin{equation}
^{(5)\!}R_{ABCD}=\frac{\Lambda_5}{6} \left[\,
^{(5)\!}g_{AC}\,^{(5)\!}g_{BD}-\,^{(5)\!}g_{AD}\,^{(5)\!}g_{BC}
\right] \,.
\end{equation}
Therefore to linear order in the metric perturbations, and
enforcing the transverse and traceless conditions, the field
equations in the absence of matter are given by
\begin{equation}
\label{boxgAB} ^{(5)\!}\Box \,h_{AB} =- \frac{1}{3} \Lambda_5
\,h_{AB} \,.
\end{equation}

The tracefree condition, in a GN gauge, requires
\begin{equation}
\label{tracefreeGN} A+3\R-k^2 E=0 \,.
\end{equation}
The transverse condition in general gives rise to four constraint
equations, which can be written, using Eq.~(\ref{tracefreeGN}), as
\bea \label{transverseGN}
-k^2 B +2\left(\dot A + 4\frac{\dot a}{a}A \right) &=& 0 \,, \\
\left\{ \label{transverseGN2}
\frac{a^2}{n^2}\left[\left(\frac{\dot n}{n}-5\frac{\dot
a}{a}\right)B-
\dot B\right] -2A -4\R \right\}_{,i} &=& 0 \,, \\
2\left(\frac{a'}{a} - \frac{n'}{n} \right) A &=& 0\,. \eea
Unless $(a/n)'=0$, i.e., unless $a$ and $n$ have the same
$y$-dependence, the five constraint equations require that the
four GN scalar metric perturbations are all identically zero.
Using the background field equation $^{(5)\!}G^4_0=0$ [see
Eq.~(\ref{5DEinstein})], this implies that it is only possible to
use the transverse and tracefree GN gauge for a separable bulk
metric, which corresponds to a Minkowski or (anti-)de Sitter brane
given in Eq.~(\ref{dSbackground}).

Thus, only in the special case of a (anti-)de Sitter or Minkowski
brane, the wave equation~(\ref{boxgAB}) gives an evolution
equation for the scalar metric perturbation:
\begin{equation}
\label{Aeom} \frac{1}{n^2}\left\{\ddot A -\left(\frac{\dot n}{n}-7
\frac{\dot a}{a} \right)\dot A \right\} + \left[ 8\left(\frac{\dot
a}{an}\right)^2 +2\left(\frac{a'}{a}\right)^2 -2 \mu^2 \right] A
+\frac{k^2}{a^2} A =  A''+  4\frac{a'}{a} A' \,,
\end{equation}
and the remaining scalar metric perturbations can be deduced from
the constraint
equations~(\ref{tracefreeGN})--(\ref{transverseGN2}). The GN gauge
choice necessarily eliminates the radion mode from the metric
perturbations. In this gauge the radion in a two-brane system must
be described instead as a relative perturbation of the coordinate
position of the branes.

\section{Bulk gravitons with a de Sitter brane}

\subsection{Separable background}

In order to solve for the $y$-dependence of the bulk gravitons and
to study the time-dependence of the perturbations on the brane, we
will consider the special case of a de Sitter brane (with constant
Hubble rate $H$, energy density $\rho$ and pressure $-\rho$) in an
AdS bulk, which gives a separable form for the bulk
metric~\cite{Kaloper},
\begin{eqnarray}
 \label{dSbackground}
&& ds^2 = \A^2(y) \left[ -dt^2 + a_o^2(t) d\vec{x}\,^2 \right] +
 dy^2 \,,\\
&& a_o(t) = \exp {Ht} \,,~~
 \A(y) =
 {H\over\mu}\sinh\mu(y_{\rm h}- |y|)\,,\label{Ay}
\end{eqnarray}
where $y=\pm y_{\rm h}$ are Cauchy horizons, with
\begin{equation}
y_{\rm h} ={1\over\mu}\coth^{-1} \left(1+ {\rho\over\lambda}
\right)\,.
\end{equation}

Any constant-$y$ hypersurface corresponds to an exponentially
expanding de Sitter slice for $\rho>0$, giving a dS$_4$ slicing of
AdS$_5$. The original RS solution~\cite{RS99} with Minkowski
spacetime on the brane (M$_4$ slicing of AdS$_5$) is recovered in
the limit $\rho/\lambda\to0$, when $\A\to \exp(-\mu|y|)$ and
$y_{\rm h} \to\infty$.  At very high energies, $\rho\gg\lambda$,
deviations from the RS solution will be significant. The junction
conditions in Eq.~(\ref{junction_back}) require that
$p=-\rho=\,$constant. This will be a good approximation to a
potential-dominated scalar field rolling slowly down a
sufficiently flat potential~\cite{MWBH99}.

It is often useful to work in terms of the conformal
bulk-coordinate $z=\int dy/\A(y)$:
\begin{equation}
 \label{defz}
z = {\rm sgn}(y) H^{-1}
\ln\left[\coth{\textstyle{1\over2}}\mu(y_{\rm h}-|y|) \right]\,.
\end{equation}
The Cauchy horizon is now at $|z|=\infty$, and the brane is
located at $z=\pm z_{\rm b}$, with
\begin{equation}\label{zb}
z_{\rm b} ={1\over  H}\sinh^{-1}{H\over\mu}\,.
\end{equation}
The line element, Eq.~(\ref{dSbackground}), becomes
\begin{eqnarray}
ds^{2} &=& \A^2(z) \left[ -dt^2 + dz^2 +{\rm e}^{2Ht}d\vec{x}\,^2
\right]\,, \label{dSmetric}\\
\A(z) &=& { H\over \mu \sinh H|z|}\,. \label{A}
\end{eqnarray}
In the RS limit, $\rho\to0$ and $H\to0$, so that
$\A\to[1+\mu(|z|-z_{\rm b})]^{-1}$.

\subsection{Master variable}

In the dS$_4$ slicing of AdS$_5$, the master variable wave
equation~(\ref{scalarmastereom}) reduces to
\begin{equation}
\frac{1}{\A^2} \left( - \ddot\Omega + 3H\dot\Omega \right)
 + \Omega'' - 2\frac{\A'}{\A}\Omega' = \left( \frac{k^2}{a_o^2\A^2} -
 \mu^2 \right) \Omega \,.
\end{equation}
The solutions can be separated into eigenmodes of the
time-dependent equation on the brane and the bulk mode equation,
$\Omega(t,y;\vec{x})= \int d^3\vec{k}\, dm\, \alpha_m(t) \H_m(y)
e^{i\vec{k}\cdot\vec{x}}$, where
\begin{eqnarray}
\ddot{\alpha}_m -3H\dot{\alpha}_m+\left[ m^2+{k^2\over
a_o^2}\right] \alpha_m &=&0\,, \label{varphieom}\\
\H_m''-2{\A'\over\A}\H_m' + \left[{m^2\over \A^2} + \mu^2 \right]
\H_m &=& 0\,. \label{bulkmodeeq}
\end{eqnarray}
Note that the Hubble damping term $-3H\dot\alpha_m$ has the
``wrong sign'', i.e., this is not the standard wave equation for a
scalar field in 4D. We recover the RS solutions in the limit
$H\to0$, $\rho\to0$, in which case $\varphi_m=\exp(\pm { i}\omega
t)$, with $\omega^2=k^2+m^2$, and $\H_m$ can be given in terms of
Bessel functions~\cite{RS99}.

If we write $\alpha_m=a_o^2\varphi_m$ and use conformal time
$\eta=-1/(a_oH)$, Eq.~(\ref{varphieom}) can be rewritten as
\begin{equation}
{d^2 \varphi_m \over d\eta^2}
 + \left[ k^2 - {2-(m^2/H^2) \over \eta^2} \right] \varphi_m = 0 \,.
\end{equation}
This is the same form of the time-dependent mode equation commonly
given for a massive scalar field in 4D de Sitter spacetime. The
general solution is
\begin{equation}
\varphi_m(\eta;\vec{k}) = \sqrt{-k\eta}\, Z_\nu(-k\eta)\,, ~~
\nu^2={9\over4}-{m^2\over H^2}\,, \label{varphisol}
\end{equation}
where $Z_\nu$ is a linear combination of Bessel functions of order
$\nu$. The solutions oscillate at early-times/small-scales for all
$m$, with an approximately constant amplitude while they remain
within the Hubble radius ($k\gg a_oH$).  `Heavy modes', with
$m>{3\over2}H$, continue to oscillate as they are stretched to
super-Hubble  scales, but their amplitude rapidly decays away,
$|\varphi_m^2|\propto a_o^{-3}$. But for `light modes' with
$m<{3\over2}H$, the perturbations become over-damped at
late-times/large-scales ($k\ll a_oH$), and decay more slowly:
$|\varphi_m^2|\propto a_o^{2\nu-3}$.

Defining $\Psi_m\equiv \A^{-3/2}\H_m$, it is possible to rewrite
the off-brane equation~(\ref{bulkmodeeq}) in Schr\"odinger-like
form,
\begin{equation}
\label{SE}
 {d^2\Psi_m\over dz^2} - V\Psi_m =-m^2 \Psi_m \,,
\end{equation}
where
\begin{eqnarray}
V(z)=-{{1\over4}}\mu^2\A^2(z)+{{9\over4}}H^2 = - {H^2 \over
4\sinh^2(Hz)} + {{9\over4}}H^2 \,.
\end{eqnarray}
For $z\to\infty$ we have $V\to {9\over4}H^2$ and we have a
continuum of massive modes above the mass gap~\cite{GS} $m^2>
{9\over4}H^2$ which become oscillating plane waves as
$z\to\infty$. The time-evolution of the mode functions for these
heavy modes, Eq.~(\ref{varphisol}), shows that they remain
underdamped at late times, i.e., the continuum of massive modes is
not excited by de Sitter inflation on the brane, as has previously
been noted for vector~\cite{BMWvector} and tensor~\cite{LMW}
modes.

The general solution to the Schr\"odinger equation~(\ref{SE}) is
(for $y \geq 0$)~\cite{Legendre}
\begin{equation}
\Psi_{m}= \left[ \sinh \mu (y_h-y) \right]^{-1/2} W_{\nu-1/2}
(\coth \mu (y_h- y)) =\left(\sinh Hz \right)^{{1}/{2}} W_{\nu-1/2}
(\cosh H z),
\end{equation}
where $W_{\alpha}$ is a linear combination of Legendre polynomials
of order $\alpha$. In a single-brane model, the normalization of
the solution is determined by the condition
\begin{equation}
\int^{\infty}_{z_b} \vert \Psi_m \vert^2 dz = \int^{y_h}_0  N^{-4}
\vert u_m \vert^2 dy< \infty. \label{normalization}
\end{equation}
The general solution will be non-normalizable for light massive
modes with $m^2\leq {9\over4}H^2$ (which diverge as $z\to\infty$).
Modes with $m^2\leq {9\over4}H^2$ are only normalizable if the
boundary conditions at $z=z_b$ allow us to kill the divergent part
of the solution at $z\to\infty$. There are no such modes for a
single vacuum de Sitter brane~\cite{fk}, but there is one
normalizable light mode when a second de Sitter brane is
present~\cite{GenSasaki}.

\subsubsection{Boundary conditions at the brane}

In terms of the master variable, $\Omega$, in the AdS background
with dS brane, the boundary
conditions~(\ref{perturbjunction})--(\ref{perturbjuncTT}) require
\begin{eqnarray}
\left(\ddot{\Omega}'-\frac{\A'}{\A}\ddot{\Omega} \right) +2 H
\left(\dot{\Omega}'-\frac{\A'}{\A} \dot{\Omega} \right) &=&
\kappa_5^2 a_o \delta P
\,,\\
\dot\Omega' - \frac{\A'}{\A} \dot\Omega &=& \kappa_5^2 a_o \delta
q
\,,\\
-3 H \left( \dot\Omega' - \frac{\A'}{\A} \dot\Omega \right) -
\frac{k^2}{a^2} \left( \Omega' - \frac{\A'}{\A} \Omega \right)
 &=&  \kappa_5^2 a_o \delta\rho \,.
\end{eqnarray}
For a vacuum brane, $\delta T_\mu^\nu=0$, these reduce to a single
boundary condition on the master variable,
\begin{equation}
\label{vacuumboundary} \Omega' = \frac{\A'}{\A} \Omega \,.
\end{equation}

\subsubsection{Radion mode}

The vacuum boundary condition~(\ref{vacuumboundary}) is trivially
satisfied for any $z$ by the mode
\begin{equation}
 u_{\rm r} \propto \A \,,~~ m^2=m_{\rm r}^2=2H^2\,,
\end{equation}
which is a solution of the bulk mode equation~(\ref{bulkmodeeq}).
The Schr\"odinger wave function $\Psi_{\rm r}\propto \A^{-1/2}$
diverges as $z\to\infty$ (where $\A\to0$) so this mode is
non-normalizable in the single-brane model. However in a
stationary two-brane model, where the second brane is at any fixed
$z_2>z_b$, this mode is normalizable~\cite{GenSasaki} and
automatically obeys the boundary condition~(\ref{vacuumboundary})
for any $z_2$.

We identify this mode as the ``radion'', which exists in the
two-brane model but is non-normalizable for a single brane. The
time dependence of this mode is given by Eq.~(\ref{varphisol})
with $\nu={1\over2}$,
\begin{equation}
\varphi_{\rm r}(\eta,k) = \sqrt{-k\eta}\, Z_{1/2}(-k\eta)\,,
\end{equation}
and hence the master variable on the brane on large scales or at
late times, for which $k/a_oH\ll1$, is
\begin{equation}
\Omega_{\rm r} \approx C_1 a_o^2 + C_2 a_o \,.
\end{equation}

The physical effect of the radion on the brane-world can be
interpreted as an effective energy-momentum perturbation. The
perturbed 5D Weyl tensor is felt on the brane through its
projection $\delta E_{\mu\nu}$, which has an effective energy and
momentum density on the brane, given in terms of the master
variable by~\cite{Deffayet}
\begin{eqnarray}
\label{Weylrho}
\kappa_4^2 {\delta\rho}_E &=& \frac{k^4}{3a_o^5} \Omega \,,\\
\kappa_4^2 {\delta q}_E &=& \frac{k^2}{3a_o^2} \left(
  \frac{\Omega}{a_o} \right)^{\displaystyle\cdot}
 \,,\\
\label{Weylpi} \kappa_4^2 {\delta\pi}_E &=& \frac{1}{2a_o^3}
\left( \ddot\Omega -
  H\dot\Omega + \frac{k^2}{3a_o^2}\Omega \right) \,.
\end{eqnarray}
The time dependence of the radion mode on large scales gives the
physical effect, from Eqs.~(\ref{Weylrho})--(\ref{Weylpi}), as
\begin{eqnarray}
\kappa_4^2 {\delta\rho}_{E\,\rm r} &\approx& H^4 \left[
\frac{C_1}{3} a_o \left( \frac{k}{a_o H} \right)^4
   + \frac{C_2}{3} \left( \frac{k}{a_o H} \right)^4 \right] \,,
\label{Weylmaster0} \\
\kappa_4^2 {\delta q}_{E\,\rm r} &\approx& H^3 \left[
\frac{C_1}{3} a_o \left( \frac{k}{a_o H} \right)^2 +\frac{C_2}{9}
\left( \frac{k}{a_o H} \right)^4 \right]
  \\
\kappa_4^2 {\delta\pi}_{E\,\rm r} &\approx& H^2 \left[
\frac{C_1}{a_o} +{C_2\over45 a_o^2} \left( \frac{k}{a_o H}
\right)^4 \right]. \label{Weylmaster}
\end{eqnarray}
In order to derive the contribution from the decaying mode, we
expanded the solution for $\Omega_{\rm r}$ as
\begin{equation}
\Omega_{\rm r} =C_1 a_o^2 + C_2 a_0 \left[ 1 - \frac{1}{6} \left(
\frac{k }{a_o H} \right)^2 + \frac{1}{120} \left( \frac{k}{a_oH}
\right)^4 \right].
\end{equation}
The decaying mode corresponds to dark radiation with isotropic
pressure ${\delta P}_E={1\over3}{\delta\rho}_E$~\cite{Yoshiguchi},
but the dominant mode is supported by a non-negligible momentum
density on large scales, driven by the anisotropic pressure
exerted by the radion field~\cite{GenSasaki}.

\subsection{Gaussian normal gauge}

In the 5D longitudinal gauge, the radion is encoded as a discrete
mode in the bulk metric perturbations. In the GN coordinates, the
radion must be described instead as a relative perturbation of the
coordinate position, i.e. the brane-bending scalar $\xi$.  In this
subsection, we show that these two descriptions are equivalent.

In the dS$_4$ slicing of AdS$_5$, Eq.~(\ref{Aeom}) reduces to
\begin{equation}
-\frac{1}{\A^2} \left(\ddot{A}+7 H \dot{A} +10 H^2 A
+\frac{k^2}{a_o^2} A \right)+A''+4 \frac{\A'}{\A} A'=0.
\end{equation}
This wave equation can be separated via $A(t,y;\vec{x})= \int
d^3\vec{k}\, dm\, f_m(t) g_m(y) e^{i\vec{k}.\vec{x}}$, as
\begin{eqnarray}
\ddot{f}_m+7 H \dot{f}_m+10 H^2 f_m+\left[ m^2+ \frac{k^2}{a^2}
f_m
\right]&=& 0,  \\
\label{gm} g_m''+4 \frac{\A'}{\A} g_m' +\frac{m^2}{\A^2} g_m &=&
0.
\end{eqnarray}
Defining $\Phi_m\equiv \A^{3/2}g_m$, we can rewrite the off-brane
equation in the same Schr\"odinger-like form as Eq.~(\ref{SE}),
but with potential
\begin{eqnarray}
V(z)&=&\frac{15}{4} \frac{H^2}{\sinh^2 (H z)} + \frac{9}{4} H^2.
\end{eqnarray}
This is the same effective potential as for tensor
perturbations~\cite{LMW}. Again, modes with $0< m^2 \leq {9\over4}
H^2$ are not normalizable in a single-brane model unless the
boundary conditions at $z=z_{\rm b}$ kill off the divergent part
of the solution at $z \to \infty$.

\subsubsection{Boundary conditions}

In the (general) GN gauge, the boundary conditions for metric
perturbations, Eqs.~(\ref{perturbjunction})--(\ref{perturbK}), for
the vacuum brane include a contribution from the brane-bending
scalar $\xi$:
\begin{eqnarray}
\label{Aprime}
A' &=& -\ddot{\xi}+H^2 \xi,  \\
\R' &=& -H \dot{\xi} +H^2 \xi,  \\
E' &=& {1\over a_o^2} \xi.
\end{eqnarray}
The evolution equation for $\xi$ can be derived from the
$y$-derivative the traceless condition~(\ref{tracefreeGN}),
\begin{equation}
A'+3 \R'-k^2 E'=0,
\end{equation}
which yields
\begin{equation}
\ddot{\xi}+3 H \dot{\xi}-4 H^2 \xi +{k^2\over a_o^2} \xi=0.
\end{equation}
We note that the brane-bending has a tachyonic effective mass,
$m^2=-4 H^2$, for a de Sitter brane~\cite{GenSasaki,ChackoFox}.

\subsubsection{Radion mode}

It is possible to find a particular solution for $A$ supported by
the brane-bending scalar~\cite{Minamitsuji}
\begin{equation}
A(t,y)=F(y)\left[-\ddot{\xi}(t)+H^2 \xi(t)\right],
\label{solutionA}
\end{equation}
where $F$ obeys
\begin{equation}
F''+4 \frac{\A'}{\A} F' +\frac{2 H^2}{\A^2} F =0.
\end{equation}
Comparing this bulk equation with Eq.~(\ref{gm}), we see that the
radion supports a discrete bulk mode with $m^2 =2 H^2$, and $F(y)$
is given by
\begin{equation}
F(y)=D_1 \coth \mu(y_h-\vert y \vert) +D_2 \left[1+ \coth^2
\mu(y_h-\vert y \vert) \right],
\end{equation}
where $D_1$ and $D_2$ are integration constants. The boundary
condition~(\ref{Aprime}) requires $F'(0)=1$ and hence gives one
relation between $D_1$ and $D_2$. From the constraint equations we
get the solutions for other metric perturbations:
\begin{eqnarray}
B &=& {2F(y)\over a_o^2}\left[\dot{\xi}-H \xi \right],
\label{solutionGN0}
\\
\R &=& F(y) H \left[-\dot{\xi}+H \xi \right],  \\
E &=& {F(y)\over a_o^2}  \xi\,. \label{solutionGN}
\end{eqnarray}
This mode is not normalizable in a single-brane model. In a static
two-brane model, it becomes normalizable, and we need to consider
the bending of the second brane, $\xi_2$. Then we replace $\xi$ by
$\xi-\xi_2$ in the final result and $\xi-\xi_2$ satisfies the same
4D wave equation as $\xi$. As expected, the radion in a two-brane
system is described as a relative perturbation of the coordinate
position of the branes $\xi-\xi_2$. The radion supports a discrete
bulk mode with $m^2=2 H^2$.

\subsubsection{Projected Weyl tensor}

The equivalence of the two descriptions of the radion can be shown
by evaluating the projected Weyl tensor. The effective
energy-momentum tensor of the projected Weyl tensor is simply
related to normal derivatives of the GN metric
perturbations~\cite{BMW},
\begin{eqnarray}
\kappa_4^2 {\delta\rho}_E &=& - \left( A'' + 2\frac{\A'}{\A}
  A' \right) \,,\\
\kappa_4^2 {\delta q}_E &=& - \frac{a_o^2}{2} \left( B'' + 2
  \frac{\A'}{\A} B' \right) \,,\\
\kappa_4^2 {\delta\pi}_E &=& E'' + 2\frac{\A'}{\A} E' \,.
\end{eqnarray}
The solution for $\xi$ on large scales is
\begin{equation}
\xi=c_1 a_o \left[1+\frac{1}{6} \left(\frac{k}{a_o H}\right)^2 +
\frac{1}{24}\left(\frac{k}{a_o H}\right)^4   \right] + {c_2\over
a_o^4}.
\end{equation}
Then, from the solutions for the metric perturbations
Eqs.~(\ref{solutionA}), (\ref{solutionGN0})--(\ref{solutionGN}),
we can evaluate the projected Weyl tensor,
\begin{eqnarray}
\kappa_4^2 {\delta\rho}_E &=& \frac{2 \mu^2 H^2}{\sinh^4 \mu
y_h}D_2 \left[{c_1\over 3}a_o \left(\frac{k}{a_o H}\right)^4
+15 {c_2\over a_o^4} \right], \\
\kappa_4^2 {\delta q}_E &=& \frac{2 \mu^2 H}{\sinh^4 \mu y_h} D_2
\left[{c_1\over3}a_o \left(\frac{k}{a_o H}\right)^2 + 5{c_2 \over
a_o^4} \right],
\\
\kappa_4^2{\delta\pi}_E &=& \frac{2 \mu^2}{\sinh^4 \mu y_h} D_2
\left[{c_1\over a_o} +{c_2\over a_o^6} \right].
\end{eqnarray}
These agree with the results obtained using the master variable,
Eqs.~(\ref{Weylmaster0})--(\ref{Weylmaster}).

\section{Scalar field on the brane}

The simplest dynamical model of inflation on the brane involves a
scalar field confined to the brane, which obeys the standard 4D
wave equation on the brane:
\begin{equation}
\label{phieom} \Box\phi = \frac{dV}{d\phi} \,.
\end{equation}
In the original computation~\cite{MWBH99} of the spectrum of
scalar perturbations generated by such a slow-roll brane inflation
scenario, it has been assumed that, since the scalar field in this
scenario is intrinsically 4D, the usual formula for the quantum
fluctuations of a 4D scalar field should apply, giving
$\delta\phi\sim H/(2\pi)$ at Hubble crossing. This should be valid
for linear perturbations of a massless scalar field in de Sitter
spacetime where the perturbations in the energy-momentum tensor
are only second-order in the field fluctuations. On scales much
larger than the Hubble radius at the end of slow-roll inflation,
one can then calculate the curvature perturbation on
uniform-density hypersurfaces which should be conserved, so long
as energy is conserved on the brane, for adiabatic perturbations
\cite{WMLL,LMSW}. The only difference from the standard
inflationary calculation of field fluctuations then comes from the
fact that the background Hubble rate $H$ is governed by the
modified Friedmann equation~\cite{BDL}.

However, one would like to check whether 5D effects could spoil
this reasoning. In particular the inflaton perturbations are
linked with the 5D metric perturbations at first-order in a
slow-roll expansion via the junction conditions. In this section
we investigate the nature of the scalar metric perturbations that
are produced by field fluctuations on the brane.

\subsection{Bulk scalar modes}

In previous sections, we have seen that in a single-brane model,
there are no light modes for a vacuum brane, and that in a
two-brane model the only light mode (with $m^2< {9\over4}H^2$) is
the radion mode. In this section we include matter perturbations
on the brane and show, using the master variable to describe bulk
metric perturbations, that the matter perturbations support an
infinite ladder of normalizable modes.

For the matter, we consider a 4D inflaton scalar field $\phi$ with
potential $V(\phi)$. Scalar field perturbations have vanishing
anisotropic stress at linear order, $\delta\pi=0$ in
Eq.~(\ref{perturbjuncTT}), and hence the boundary condition at the
brane for the off-diagonal part of the extrinsic curvature in
Eq.~(\ref{perturbK}) requires the brane position to be unperturbed
($\xi=0$) in the 5D longitudinal gauge. In this case the 5D
longitudinal gauge coincides with the 4D longitudinal gauge on the
brane. The remaining boundary conditions for the master variable
$\Omega$ can then be written in the general form~\cite{koyama04}
\begin{eqnarray}
a_o \kappa_5^2 \delta \rho &=& -\frac{k^2}{a^2}
\left(\Omega'-\frac{a'}{a} \Omega \right) - 3 \frac{\dot{a}}{a}
\left( \dot{\Omega}'- \frac{n'}{n} \dot{\Omega} \right),
\label{bomega0}
 \\
a_o \kappa_5^2 \delta q &=&
- \left(\dot{\Omega}' - \frac{n'}{n} \dot{\Omega}  \right),\\
a_o \kappa_5^2 \delta P &=& \ddot{\Omega}'-\frac{a'}{a}
\ddot{\Omega} +2 \frac{\dot{a}}{a} \left(
\dot{\Omega}'-\frac{n'}{n} \dot{\Omega} \right) + \left\{4
\frac{\dot{a}}{a} \left(\frac{a'}{a}-\frac{n'}{n} \right) + 2
\left(\frac{\dot{a}}{a}\right)' -
\left(\frac{\dot{n}}{{n}}\right)'  \right\} \dot{\Omega}
 \nonumber\\
&&~{}- \frac{2}{3} \left( \frac{a'}{a}-\frac{n'}{n} \right)
\frac{k^2}{a^2} \Omega +\mu^2 \left(\frac{a'}{a}-\frac{n'}{n}
\right) \Omega -\left(\frac{a'}{a}-\frac{n'}{n}\right)
\left(2 \frac{a'}{a}- \frac{n'}{n} \right) \Omega', \nonumber\\
\label{bomega}
\end{eqnarray}
where the brane matter perturbations on the left-hand sides will
be expressed in terms of the scalar field perturbation
$\delta\phi$ and of the induced metric perturbations in the 4D
longitudinal gauge.

For $dV/d\phi\neq0$, the brane is no longer strictly de Sitter,
but in order to make the problem tractable, we impose two
approximations. The first is to assume zeroth-order slow-roll for
the background, which means that in practice we consider the
background as a strict de Sitter brane configuration. The second
simplification is to ignore the brane metric perturbation
contributions to $\delta\rho$, $\delta P$.  In the standard 4D
calculation, this latter approximation is known to be valid in the
{\it slow-roll limit} in the 4D longitudinal gauge. In the present
case, this can be justified only in retrospect once we have done
the simplified calculation.

With these two approximations, the junction
conditions~(\ref{bomega0})--(\ref{bomega}) reduce to
\begin{eqnarray}
\left(\ddot{\Omega}'-\frac{\A'}{\A}\ddot{\Omega} \right) +2 H
\left(\dot{\Omega}'-\frac{\A'}{\A} \dot{\Omega} \right) &=&
\kappa_5^2 a_o \left[\dot{\phi} \dot{\delta \phi} -V'(\phi) \delta
\phi \right]
\,,\\
\dot\Omega' - \frac{\A'}{\A} \dot\Omega &=& \kappa_5^2 a_o
\dot{\phi} \delta \phi \,, \label{scalar}
\\
-3 H \left( \dot\Omega' - \frac{\A'}{\A} \dot\Omega \right) -
\frac{k^2}{a_o^2} \left( \Omega' - \frac{\A'}{\A} \Omega \right)
&=& \kappa_5^2 a_o \left[\dot{\phi} \dot{\delta \phi} + V'(\phi)
\delta \phi\right] \,,
\end{eqnarray}
where the contributions from the induced metric perturbations on
the right-hand sides are neglected.

Defining
\begin{equation}
{\cal F}(t)= \Omega'-\frac{\A'}{\A} \Omega,
\end{equation}
at the brane, and combining the junction conditions, we get a
single evolution equation,
\begin{equation}
\ddot{{\cal F}}-  \left(H + 2 \frac{\ddot{\phi}}{\dot{\phi}}
\right) \dot{{\cal F}} + \frac{k^2}{a_o^2} {\cal F}=0. \label{eqF}
\end{equation}
This gives the boundary condition for the time dependence of the
master variable $\Omega$. From Eq.~(\ref{scalar}) the scalar field
fluctuation $\delta \phi$ is given in terms of ${\cal F}$ by
\begin{equation}
\kappa_5^2 \delta \phi = \frac{\dot{{\cal F}}}{a_o \dot{\phi}}~.
\label{scalar2}
\end{equation}
Then it can be verified that Eq.~(\ref{eqF}) is consistent with
the equation of motion for $\delta \phi$,
\begin{equation}
\label{zerothphi} \ddot{\delta \phi}+3 H \dot{\delta \phi}+{k^2
\over a_o^2} \delta \phi + V''(\phi) \delta \phi=0\,,
\end{equation}
for an arbitrary $V(\phi)$, to lowest order (i.e., neglecting the
metric perturbations).

Assuming that $\phi$ is slow-rolling, so that
$\,|\ddot\phi/\dot\phi|\ll H\,$ in Eq.~(\ref{eqF}), the solution
for ${\cal F}$ is
\begin{equation}
{\cal F}(\eta)=C_1 \frac{ \cos (-k \eta)}{-k \eta} + C_2
 \frac{\sin (-k \eta)}{-k \eta}.
\label{F}
\end{equation}
This should be compared with the time evolution of $\Omega$ given
by each of the mode functions $\alpha_m$ in Eq.~(\ref{varphisol}).
One might expect that the boundary condition can be satisfied by
summing up mode functions only with positive $m^2$. However, it
turns out that this is not possible, and negative-$m^2$ modes are
unavoidable. We use the formulas for summation of Bessel
functions,
\begin{eqnarray}
\sum^{\infty}_{\ell=0} (-1)^{\ell} \left( 2 \ell+\frac{3}{2}
\right)z^{-3/2} J_{2\ell + 3/2}(z)
&=& \sqrt{\frac{1}{2 \pi}}\, \frac{\sin z}{z},  \\
\sum^{\infty}_{\ell=0} (-1)^{\ell} \left( 2 \ell+\frac{1}{2}
\right)z^{-3/2} J_{2\ell + 1/2}(z) &=& \sqrt{\frac{1}{2 \pi}}\,
\frac{\cos z}{z}\,.
\end{eqnarray}
These show that an infinite sum of mode functions
\begin{equation}
\alpha_m=(-k\eta)^{-3/2}J_\nu(-k\eta) \,,
 \quad {\rm where}\ \nu^2={9\over 4}-{m^2\over H^2}\,,
\end{equation}
can satisfy the boundary condition imposed on ${\cal F}$, where
the spectrum of KK modes is given by
\begin{eqnarray}
\frac{m^2}{H^2}&=&-2(2\ell-1)(\ell+1) \quad \mbox{for} \ C_1,  \\
\frac{m^2}{H^2}&=&-2 \ell (2\ell+3) \quad \mbox{for} \ C_2.
\end{eqnarray}
These modes include an infinite ladder of tachyonic modes with
$m^2<0$. However, the boundary condition requires us to include
only the decaying solution for these tachyonic modes. The
dangerous growing mode solution is excluded once the junction
condition is imposed. Thus there is no instability.

We should choose the solution in the $y$-direction so that the
metric perturbations remain small as $y \to y_h$ and the mode is
normalizable for a single brane. Unlike the case of the radion
mode for a vacuum brane, the test scalar field on the brane allows
us to choose only the normalizable modes. The solution for
$\Omega$ in the bulk is (for $y\geq0$)
\begin{eqnarray}
\Omega(\eta,y) &=& C_1 \sqrt{2 \pi}\, \sum_{\ell=0}^{\infty}
(-1)^{\ell}\left(2\ell+\frac{1}{2} \right) \frac{\sinh
\mu(y_h-y)\, Q_{2\ell}(\coth \mu(y_h-y))} {\mu \,
Q^{1}_{2\ell}(\coth \mu y_h)}\, (-k \eta)^{-3/2} J_{2\ell+1/2}(-k
\eta)  \nonumber\\ \nonumber\\ && {}+ C_2 \sqrt{2 \pi}\,
\sum_{\ell=0}^{\infty} (-1)^{\ell}\left(2\ell+\frac{3}{2} \right)
\frac{\sinh \mu(y_h-y)\,  Q_{2\ell+1}(\coth \mu(y_h-y))} {\mu \,
Q^{1}_{2\ell+1}(\coth \mu y_h)} \,(-k \eta)^{-3/2}
J_{2\ell+3/2}(-k \eta)\,, \label{solutionomega}\\ \nonumber
\end{eqnarray}
where $Q_{\alpha}$ is a Legendre polynomial of the second kind and
$Q^{\alpha}_{\beta}$ is an associated Legendre function of the
second kind. Using the asymptotic behavior of $Q_{n}(z)$,
\begin{equation}
Q_{n}(z) \to \sqrt{\pi}\, \frac{\Gamma(n+1)}{\Gamma \left(n+
\frac{3}{2}\right)}\, (2z)^{-n-1} \quad (\mbox{for} \quad z \to
\infty),
\end{equation}
we can check that this solution is normalizable [see
Eq.~(\ref{normalization})]. Thus if there is a matter perturbation
on the brane, normalizable discrete modes are supported in a
single-brane model. In the following discussion, we concentrate on
a single-brane model.

On large scales, $k \eta \to 0$, the Bessel function behaves as
$J_{\nu}(-k \eta) \propto a_o^{-\nu}$, so that the mode with
smallest $\nu$, i.e. the mode with $m^2=2 H^2$, gives the dominant
contribution in the $C_1$ mode. Thus on large scales, the solution
for $\Omega$ is given by the $m^2=2H^2$ mode,
\begin{equation}
\Omega_{m^2=2H^2}= C_1 a_o \,\mu (y_h-y) \sinh \mu (y_h-y) .
\label{H2}
\end{equation}
Then we can determine the solution for metric perturbations,
\begin{equation}
\R=-A=\frac{1}{2} C_1 \mu^2 \cosh \mu y_h.
\end{equation}
The scalar field perturbation is also given in terms of $C_1$ from
Eq.~(\ref{scalar2}),
\begin{equation}
\kappa_5^2 \delta \phi = -\frac{H}{\dot{\phi}} C_1 \mu\sinh \mu
y_h\,.
\end{equation}
The relation between scalar field and metric perturbations,
\begin{equation}
\R=-\frac{\dot{\phi}}{2 H} \kappa_{4,{\rm eff}}^2 \delta \phi
\quad \mbox{where}~ \kappa_{4,{\rm eff}}^2= \kappa_5^2 \mu \left[1
+ \left( \frac{H}{\mu} \right)^2 \right]^{1/2} ,
\end{equation}
is the same as the standard 4D result, except for the high-energy
correction of the 4D Newton constant.

\subsection{Metric backreaction}

We now investigate the corrections to the evolution of scalar
field fluctuations which come from the effect of the metric
perturbations that the field fluctuations themselves induce on the
brane.

We expand the scalar field perturbations in terms of slow-roll
parameters,
\begin{equation}
\label{srexpand} \delta \phi= \delta \phi_0+\delta \phi_1 +... \,,
\end{equation}
where the zeroth-order order solution corresponds to the solution
to Eq.~(\ref{zerothphi}). The first-order equation can be derived
from the scalar field equation of motion~(\ref{phieom}),
\begin{equation}
\ddot{\delta \phi}_1 + 3H \dot{\delta \phi}_1 +\frac{k^2}{a_o^2}
\delta \phi_1 =-V'' \delta \phi_0 - 3 \dot{\phi} \dot{\R} +
\dot{\phi} \dot{A}-2 V' A\,,
\end{equation}
where we have included the scalar metric perturbations in the 4D
longitudinal gauge induced by the zeroth-order field fluctuations,
$\delta\phi_0$.

The perturbed effective Einstein equations~(\ref{EE}) on the brane
are given by
\begin{eqnarray}
\frac{\kappa_{4,{\rm eff}}^2}{2} \left(\dot{\phi} \dot{\delta
\phi}_0 +V' \delta \phi_0 \right)
+\frac{\kappa_4^2}{2}\delta\rho_E
&=&3 H \dot{\R}-3 H^2 A + \frac{k^2}{a_o^2} \R, \\
\frac{\kappa_{4,{\rm eff}}^2}{2} \left(\dot{\phi} \dot{\delta
\phi}_0 -V' \delta \phi_0 \right) +\frac{\kappa_4^2}{6}
{\delta\rho}_E &=& -\ddot{\R}-3H \dot{\R}+H \dot{A}+3 H^2 A
-\frac{1}{3} \frac{k^2}{a_o^2} (\R+A), \\
\kappa_4^2 {\delta\pi}_E &=& -\frac{1}{a_o^2}(\R+A).
\end{eqnarray}
In order to evaluate the effect of metric perturbations, it is
useful to use the Mukhanov-Sasaki variable,
\begin{equation}
\label{Q} {\cal Q} =\delta \phi-\frac{\dot{\phi}}{H} \R \,.
\end{equation}
In terms of our slow-roll expansion, Eq.~(\ref{srexpand}), we have
${\cal Q}_0 =\delta \phi_0$ and ${\cal Q}_1=\delta
\phi_1-(\dot{\phi}/H) \R$. Then following~\cite{Koyama}, we use
the effective Einstein equations to derive the equation for ${\cal
Q}$,
\begin{equation}\label{q}
\ddot{{\cal Q}}_{1}+3 H \dot{{\cal Q}}_{1} + \frac{k^2}{a_o^2}
{\cal Q}_{1} =-V'' {\cal Q}_{0}- 6 \dot{H} {\cal Q}_{0} + {\cal
J},
\end{equation}
where
\begin{eqnarray}
{\cal J} &=& -\frac{\kappa_4^2 \dot{\phi}}{3 H} \left(k^2
{\delta\pi}_E + \delta\rho_E \right)
\\
&=&-\frac{\dot{\phi}}{H}\frac{k^2}{6 a_o^3} \left(\ddot{\Omega}-H
\dot{\Omega} + \frac{k^2}{a_o^2} \Omega \right).
\end{eqnarray}
Equation~(\ref{q}) is the same as the standard 4D equation except
for the term ${\cal J}$, which describes the corrections from the
5D bulk perturbations. We can evaluate the right-hand side using
Eq.~(\ref{solutionomega}) for $\Omega$ on the brane.

In order to evaluate ${\cal J}$, we need to handle the infinite
sum of modes. However, on large scales, it is possible to use the
$m^2=2 H^2$ mode, Eq.~(\ref{H2}), to rewrite ${\cal J}$ in terms
of ${\cal Q}_{0}$ as
\begin{equation}
{\cal J} =  \frac{\kappa_5^2 \mu \dot{\phi}^2}{9 H^2} (\mu y_h)
\frac{k^4}{a_o^4 \mu^2} {\cal Q}_{0}. \label{J}
\end{equation}
Here we should note that the leading-order time behaviour of the
$m^2=2 H^2$ mode on large scales, Eq.~(\ref{H2}), disappears in
${\cal J}$, so we need to take into account the next order
solution. Then the $m^2=-4 H^2$ mode gives a comparable
contribution, but it gives qualitatively the same contribution as
Eq.~(\ref{J}), so we neglect it. To compare this correction with
the standard correction term $-6 \dot{H} {\cal Q}_{0}$, we use the
background equation
\begin{equation}
\dot{H}=-\frac{1}{2} \kappa_{4,{\rm eff}}^2 \dot{\phi}^2,
\end{equation}
and $\mu y_h = \sinh^{-1}(\mu/H)$, to evaluate the ratio of these
two corrections:
\begin{equation}
\frac{{\cal J}}{\dot{H} {\cal Q}_{0}} \sim \frac{k^4}{a_o^4 \mu^2
H^2} \sinh^{-1} \frac{\mu}{H}\left[1+ \left(\frac{H}{\mu}\right)^2
\right]^{-1/2} .
\end{equation}
At low energies, $H/\mu \ll 1$, this ratio is very small on
super-Hubble scales. Even at high energies, $H/\mu \gg 1$, the
ratio is suppressed on super-Hubble scales,
\begin{equation}
\frac{{\cal J}}{\dot{H} {\cal Q}_{0}} \sim \frac{k^4}{a_o^4 H^4}.
\end{equation}

Thus we conclude that the corrections that come from bulk metric
perturbations are always small compared with the corrections to
the de Sitter geometry, i.e., $\dot{H}/H^2$ corrections, on
super-Hubble scales.

But on sub-Hubble scales, the situation changes significantly. The
correction ${\cal J}$ becomes significant, and an infinite number
of modes in $\Omega$ should be taken into account, because all
modes become comparable. This indicates that the quantum theory of
the Mukhanov-Sasaki variable on small scales is quite different
from the standard 4D results that take into account slow-roll
corrections. This is in line with the expectation that high-energy
particles on the brane can couple to massive bulk gravitons and
will be sensitive to the higher-dimensional geometry.

\section{Conclusions}
\label{conc}

In this paper, we investigated bulk scalar metric perturbations
about a de Sitter brane. In the absence of matter perturbations,
we have confirmed that there are no normalizable light modes (with
$m^2< {9\over4}H^2$) for a single brane, but in the presence of a
second brane there is a normalizable ``radion'' mode.

In the 5D longitudinal gauge the coordinate positions of vacuum
branes are unperturbed and the radion mode appears as a discrete
bulk mode with $m^2=2 H^2$. In a Gaussian normal gauge, with
transverse-tracefree condition, the radion appears as a relative
perturbation of the coordinate position of the two branes. This
``brane-bending'' mode obeys a canonical 4D wave equation with a
tachyonic effective mass, $m^2=-4H^2$, as reported in previous
analyses~\cite{GenSasaki,ChackoFox,ckp}. The brane-bending
supports a discrete bulk mode in the GN gauge~\cite{Minamitsuji},
which again has $m^2=2 H^2$.

We have shown the equivalence of the descriptions in the two
different gauges by evaluating the projected Weyl tensor on the
brane. The radion appears as an ``instability'' for two de Sitter
branes~\cite{GenSasaki,ChackoFox,ckp}, but the effect of the
radion ``instability'' on the brane measured by the projected Weyl
tensor becomes small on large scales~\cite{GenSasaki2}.

We then considered the bulk metric perturbations excited by scalar
field perturbations on a {\it single} de Sitter brane.  The $m^2=2
H^2$ mode together with the zero-mode and an infinite ladder of
discrete tachyonic modes in the 5D longitudinal gauge, become
normalizable. Nonetheless the boundary condition requires the
self-consistent 4-dimensional evolution of scalar field
perturbations on the brane, so that the dangerous growing modes
are not allowed.  These normalizable discrete modes introduce
corrections to the scalar perturbations computed in an effectively
4-dimensional approach. On super-Hubble scales, the $m^2=2 H^2$
mode is dominant and the correction is smaller than the slow-roll
corrections to the de Sitter background.  Thus we have verified
that there exists a scalar curvature perturbation $\zeta$ defined
by
\begin{equation}
\zeta = \frac{H}{\dot{\phi}} {\cal Q} \,,
\end{equation}
that is constant on large scales~\cite{WMLL,LMSW}, even including
lowest-order metric backreaction at high energies. However, on
short scales, all the infinite ladder of discrete tachyonic modes
become comparable. Thus the effect of backreaction could be large.
This is consistent with the expectation that high-energy particles
on the brane can probe the higher-dimensional gravity.

In 4-dimensional gravity we can incorporate first-order metric
perturbations along with the field fluctuations in a
gauge-invariant combination, the Sasaki-Mukhanov variable,
Eq.~(\ref{Q}), which on small scales then obeys the wave equation
for a single free field in flat spacetime. In 5-dimensional
gravity the short-wavelength field fluctuations on the brane are
coupled to an infinite ladder of bulk metric perturbations when we
try to reduce it to an effective 4-dimensional theory.

A possible consequence could be the damping of the amplitude of
quantum field fluctuations on small scales during inflation on the
brane, due to the excitation of the infinite ladder of discrete
modes. To quantify this effect, we need to be able to handle the
infinite summation of modes.  This requires further investigation
and we hope to report results in a separate publication.


\[ \]
{\bf Acknowledgments:}

KK is supported by JSPS, RM by PPARC, and DW by the Royal Society.
DL acknowledges support from the France-UK Egide programme for his
visits to Portsmouth and RM and DW acknowledge British Council
support for their visits to Paris, while this work was in
progress.



\end{document}